\documentclass[referee]{raa}

\usepackage{graphicx,times}             
\usepackage{natbib}
\usepackage{amssymb,amsmath}
\bibpunct{(}{)}{;}{a}{}{,}

\usepackage[a4paper=true,dvipdfm=true,pagebackref=true]{hyperref}
\hypersetup{colorlinks = true, linkcolor = green, anchorcolor = red, citecolor = blue, filecolor = red, pagecolor = red, urlcolor = red}

\begin{document}

   \title{Simulation of the orbit and  spin period  evolution of the double pulsars  PSR J0737-3039 from their birth to coalescence  induced by the gravitational wave radiation}

   \volnopage{Vol.0 (20xx) No.0, 000--000}       
   \setcounter{page}{1}          

   \author{Peng Liu
      \inst{1}
   \and Yi-Yan Yang
      \inst{2}
   \and Jian-Wei Zhang
      \inst{3}
  \and Maria Rah
      \inst{4}
   }

   \institute{School of Physics and Electronic Engineering, Qilu Normal University, Jinan  250200, China; {\it liupeng\_mail@sina.cn}\\
         \and 
             School of Physics and Electronic Sciences, Guizhou Education University, Guiyang 550018, China\\
          \and
             Astronomy Department, Beijing Normal University, Beijing 100875, China\\
          \and 
              Research Institute for Astronomy and Astrophysics of Maragha(RIAAM)-Maragha, IRAN\\      
  \vs\no
          {\small Received~~20xx month day; accepted~~20xx~~month day}}

\abstract{ The complete orbital and sin period evolutions of the double neutron star (NS) system  PSR J0737-3039 are simulated  from the birth to coalescence, which include  the two observed radio pulsars classified as  primary NS PSR J0737-3039A and companion NS PSR J0737-3039B. By  employing   the characteristic  age of  PSR J0737-3039B to constrain the true age of the double pulsar system, the initial orbital period and  initial binary separation  are  obtained as 2.89 hrs and  $1.44 \times 10^{6}$ km,  respectively, and the coalescence age or the lifetime from the  birth to merger  of  PSR J0737-3039 is obtained to be  $1.38 \times 10^{8}$ yr.
At the last minute of coalescence, corresponding to that the gravitational wave frequency changes from 20 Hz to 1180 Hz, we present  the binary separation of PSR J0737-3039 to be from  442 km to 30 km, while the spin periods of
PSR J0737-3039A and PSR J0737-3039B  are 27.10 ms and 4.63 s, respectively. From the standard radio pulsar emission
model, before the system merged, the primary NS  could still be observed  by the radio telescope, while the companion NS has crossed the death line in the pulsar  magnetic-field versus period ($B-P$) diagram which is usually considered
 to cease the life as a pulsar.  It is for the first time that the whole life evolutionary simulation of the orbit and spin periods for double NS  system  is presented, which provides the useful information for observing the
 primary NS at the coalescence stage.
}

      \authorrunning{Peng Liu, Yi-Yan Yang, Jian-Wei Zhang \& Maria Rah }        
      \titlerunning{Simulation of the orbit and  spin period (I) }  
   
   \maketitle

\label{sect:intro}

In 1974, Taylor and Hulse discovered the first double neutron star (DNS) system PSR J1913+16, using the Arecibo telescope ~(\citealt{Hulse+Taylor+1975}), by which,  the existence of gravitational waves (GW) as a prediction of  general relativity was indirectly verified  by studying its orbital contraction ~(\citealt{Einstein+Sitzungsber+1916, Taylor+Weisberg+1982}). More than 40 years later, for the first time, LIGO and Virgo detected GW directly in the merge of  DNS system of an old elliptical galaxy,   named as GW170817. Together with a serials of discoveries of the coalescence between the stellar black holes and NSs, the GW  theory has been confirmed and the  multi-messenger astronomy is out of birth~(\citealt{ Abbott+etal+2017, Troja+etal+2017, Blanchard+etal+2017}). Therefore, the investigation of the whole evolutionary process of the DNS system from its birth to coalescence and clarifying the details of its evolution is a matter of concern to some scientists working on NS and pulsar astrophysics.

Among the 19 discovered DNS systems until now, PSR J0737-3039 is the only known DNS system in which primary and companion NSs  have been detected as pulsars ~(\citealt{Burgay+etal+2003, Lyne+etal+2004}), which can not only provide the best direct test for the correctness of general relativity, but also can be employed as a natural lab for studying  the plasma physics and strong field gravitation~(\citealt{Kramer+etal+2004,Kramer+etal+2006,Kramer+Wex+2009}).
Only PSR J0737-3039 in 19 DNSs  can provide more information to let us know the initial orbital   and   initial NS spin periods, therefore  the PSR J0737-3039 system can be studied as a complete evolutionary  sample of the DNS systems  from the birth to the merger.
PSR J0737-3039 has a very compact orbit with the orbital period of only 2.45 hrs, and the eccentricity is as small as  0.088 (approximately a circular orbit) ~(\citealt{Lyne+etal+2004}). From the DNS formation model~(\citealt{Hulse+van den Heuvel+1991, van den Heuvel+2004}),  the primary NS (PSR J0737-3039A) was the first NS to be formed,  generated by a massive star directly through a supernova explosion and experienced accretion induced  spun-up and magnetic field decay, with the spin period of 22.7 ms and the dipole magnetic field of  $6.4\times 10^9$ G~(\citealt{Lyne+etal+2004}); the companion NS (PSR J0737-3039B) was the second NS  produced by a  electron capture supernova explosion ~(\citealt{Podsiadlowski+etal+2004, Nomoto+1984}) and ultra-stripped supernova explosion two methods  ~(\citealt{Tauris+etal+2013, Tauris+etal+2015, Tauris+etal+2017}), with the spin period and the dipole magnetic field of 2773.46 ms and $6.4\times 10^9$ G, respectively~(\citealt{Lyne+etal+2004}).

As expected, the orbits of the DNSs are contracted continuously by GW radiation and eventually  go to merge,
the direct coalescence observation and calculations of which was pointed out by  Schutz  ~(\citealt{Schutz+1986}), and by   Cutler  ~(\citealt{Cutler+etal+1992}).
Now, the numerical relativity is a very powerful tool for studying DNS orbital evolution, when the DNS system enter the merger  ~(\citealt{Maione+etal+2016, Shibata+Uryu+2000}). In order to study the complete evolution of PSR J0737-3039, we deduced the variation of the orbital period given by Peters~(\citealt{Peters+1964})  to obtain a new evolution formula and applied it to simulate the orbital evolution of DNS.
Since  PSR J0737-3039B is a non-recycled  pulsar like Crab pulsar~(\citealt{Lyne+Graham-Smith+2012, Yang+etal+2017}), and the current spin period is much larger than its  initial spin period, therefore  we can use its characteristic age to approximate the true age of PSR J0737-3039 system ~(\citealt{Camilo+etal+1994,  Lorimer+etal+2007}).
Next,  we derived the initial orbital period of the system and the initial spin periods of the two pulsars, based on which the  simulating of complete evolution of the system is presented.

The paper is organized  as bellow:
In section 2,  we introduce both of the orbital evolution formula of DNS  caused by gravitational radiation, then the simulation of the orbital complete evolutions of  PSR J0737-3039 system was given  in section 3.
We simulate the entire spin period evolution of PSR J0737-3039A and PSR J0737-3039B in section 4.
Finally, we  simulate the gravitational wave frequency of the last minute before the DNS merge, and then discuss the results.

\begin{figure}
\centering
\includegraphics[scale=0.45]{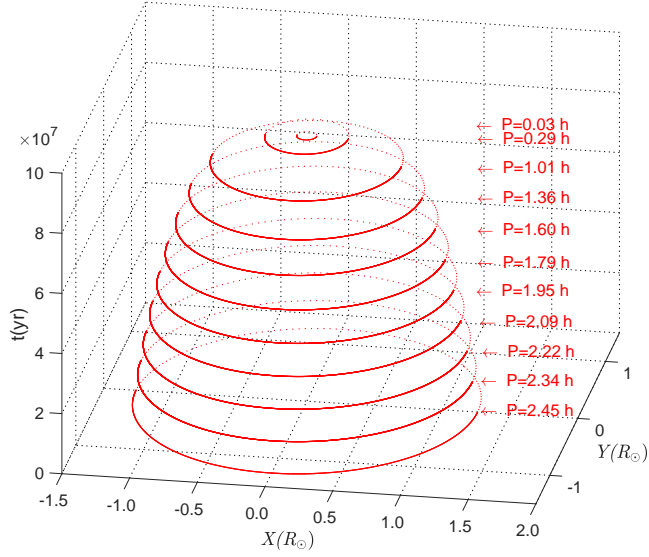}
\caption{3-D diagram  of PSR J0737-3039 evolutionary trajectory by GW radiation,
where  the vertical axis stands for the time in the units of $10^{7}$ yrs and two horizontal axes of  the orbital plane stands for the DNS orbit, with the units of radius of Sun  $R_{\odot} = 7 \times 10^{10}$ cm. }
\label{f1}
\end{figure}


\section{Orbital evolution of double neutron stars}
\label{sect:Orb}

In this section, we  derive the contraction evolution of DNS orbit by  GW radiation,  based on  the variation of the  orbit radius ($a$) of DNS system, which is  given below ~(\citealt{Peters+1964, Lyne+Graham-Smith+2012, Shapiro+Teukolsky+1983, Lightman+etal+1975, Lightman+Shapiro+1975, Ohanian+Ruffini+1994}):
\begin{equation}
\label{equ:1}
\frac{1}{a} \frac{da}{dt} = - \frac{64} {5} \frac{ G^{3}M^{2} \mu } { c^{5}a^{4} } f(e) ,
\end{equation}
where $M$ is the  total mass of DNS, expressed as $M = M_{p} + M_{c}$ with  $M_{p}$ and  $ M_{c}$  the mass of the primary  and  companion NS, respectively;
 $G$ ($c$) is the gravitational constant (speed of light); $\mu$ is the reduced mass of DNS, expressed as: $\mu = M_{p} M_{c}/(M_{p} + M_{c})$; $f(e)$ is a function of the orbital eccentricity $e$  of DNS, described in the following ~(\citealt{Peters+Mathews+1963}):

In this paper, we  only consider  the case of circular orbit $e = 0$ because the eccentricity of PSR J0737-3039 is 0.088, which arises the eccentricity function to be $f(e=0) = 1$.
\begin{equation}
\label{equ:2}
f(e) = (1 + \frac{73}{24}e^{2} + \frac{37}{96}e ^ {4}) (1-e^{2}) ^ {-\frac{7}{2}}.
\end{equation}

According to the Keplerian motion,  the relationship between the DNS orbital period and the binary separation is:
\begin{equation}
\label{equ:3}
P = ( \frac{a^{3} {4 \pi}^{2}}{GM})^\frac{1}{2},
\end{equation}
thus the orbital evolution can be also described by the orbital period equivalently.
We bring Eq.(\ref{equ:2}) and Eq.(\ref{equ:3}) into Eq.(\ref{equ:1})  and solve the evolution of  DNS binary separation with time, and obtain the following equation:

\begin{equation}
\label{equ:4}
a = a_{0}( 1 - \frac{t}{T_{gw}} )^{\frac{1}{4}},
\end{equation}
where $ a_{0}$ is  initial binary separation of DNS, and  $T_{gw}$  is a characteristic time of GW induced coalescence, expressed as:
\begin{equation}
\label{equ:5}
T_{gw}  =  \frac{5}{32} \frac{Ma_{0}^{4}} { \mu c R_{s}^{3}},
\end{equation}
with the Schwarzschild radius  $R_{s} = 2 GM/c^{2}$.

From Eq.(\ref {equ:3}) and Eq.(\ref {equ:4}), we can also deduce the relationship between orbital period and  time as follows:
\begin{equation}
\label{equ:6}
P = P_{0}( 1 - \frac{t}{T_{gw}} )^{\frac{3}{8}} ,
\end{equation}
with $P_{0}$  the initial orbital period.

Now that we have the evolutionary  equations of orbital scale and period induced by the GW radiation, the detailed simulations of double pulsars can be performed.  PSR J0737-3039 is a DNS system where two pulsars were discovered by the  pulsar search of Parkes 64 m radio telescope in 2003, lied in a distance of $1.15^{+0.22}_{-0.15}$ kpc ~(\citealt{Burgay+etal+2003, Lyne+etal+2004}). The parameters of two NSs are noted below: PSR J0737-3039A is an old, fast-spinning, recycled  pulsar, with mass  of 1.3381 $M_{\odot}$, the spin period of  22.7 ms, and the  spin period derivative is $1.76\times 10^{-18}$ s/s~(\citealt{Lyne+etal+2004}); PSR J0737-3039B is a slow-spinning, non-recycled normal pulsar with a mass of 1.2489 $M_{\odot}$, the spin period of 2773.46 ms, and the spin period derivative is $8.29\times 10^{-16}$ s/s~(\citealt{Burgay+etal+2003}). This system  has also the  shortest orbital period of 2.448 hrs among 19 pairs of DNSs, and the orbital eccentricity is as small as 0.088,  an almost circular orbit ~(\citealt{Burgay+etal+2003}). Moreover, the system presents a strong relativistic effect, orbit of which is confirmed to shrink by gravitational wave radiation during the evolution process~(\citealt{Kramer+Wex+2009, van Leeuwen+etal+2015, Abbott+etal+2009}).

Here, we simulate the orbital evolution of PSR J0737-3039 before the two NSs merge, where we assume that the two NSs  are the point particles with the nearly circular orbit.
Then, during the simulation, we regard that the DNS system begins to merge when the  orbit radius decays to the scale of two NS radii  30 km with  the conventional NS radius of 15 km.
Our simulation found that the merge  age of  PSR J0737-3039 system is $8.83 \times 10^{7}$ yr, and the maximum orbital frequency of the system before the merge  is 590 Hz, corresponding to  the GW frequency as 1180 Hz.
A schematic diagram of    orbital evolution of PSR J0737-3039 is plotted in Fig.\ref{f1}, where we draw a complete cycle of trajectory motion in every time interval $10^{7}$ yr.  in-between $0 - 8 \times10^{7}$ yr, and also include $8.8 \times10^{7}$ yr  and $8.83 \times10^{7}$ yr, and each trajectory gives the corresponding orbital period.

\begin{table}
\begin{center}
\caption[]{Parameters of double pulsar PSR J0737-3039(\cite{Lyne+etal+2004}; \cite{Burgay+etal+2003}) }\label{Tab:publ-works}
 \begin{tabular}{clclclclcl}
  \hline\noalign{\smallskip}
System                                              & $M_{p}(M_{\odot})$   &$M_{c}(M_{\odot})$   &  $P$(h)      &$P_{s}$(ms)   &d (kpc)                        & eccentricity      & $\tau$ (yr)             &$\dot{P_{s}}$(s/s)           &$B$(G)                      \\\hline\noalign{\smallskip}
J0737-3039A                                    & 1.3381(7)                  & $..........$                  & 2.448        &22.699          & $1.15^{+0.22}_{-0.15}$  & 0.088              &$2.04\times10^{8}$  &$1.76\times10^{-18}$      &$6.4\times10^{9}$        \\
J0737-3039B                                    & $..........$                  &1.2489(7)                  & $..........$     & 2773.46       & $..........$                      & $..........$          &$4.92\times10^{7}$  &$8.29\times10^{-16}$      &$1.59\times10^{12}$        \\
  \noalign{\smallskip}\hline
\end{tabular}
\end{center}
\end{table}


\section {The evolution of PSR J073-3039 system  in complete coalescence age}
\label{sect:the}

 \begin{figure*}[t]
 \centering
\includegraphics[height=5cm]{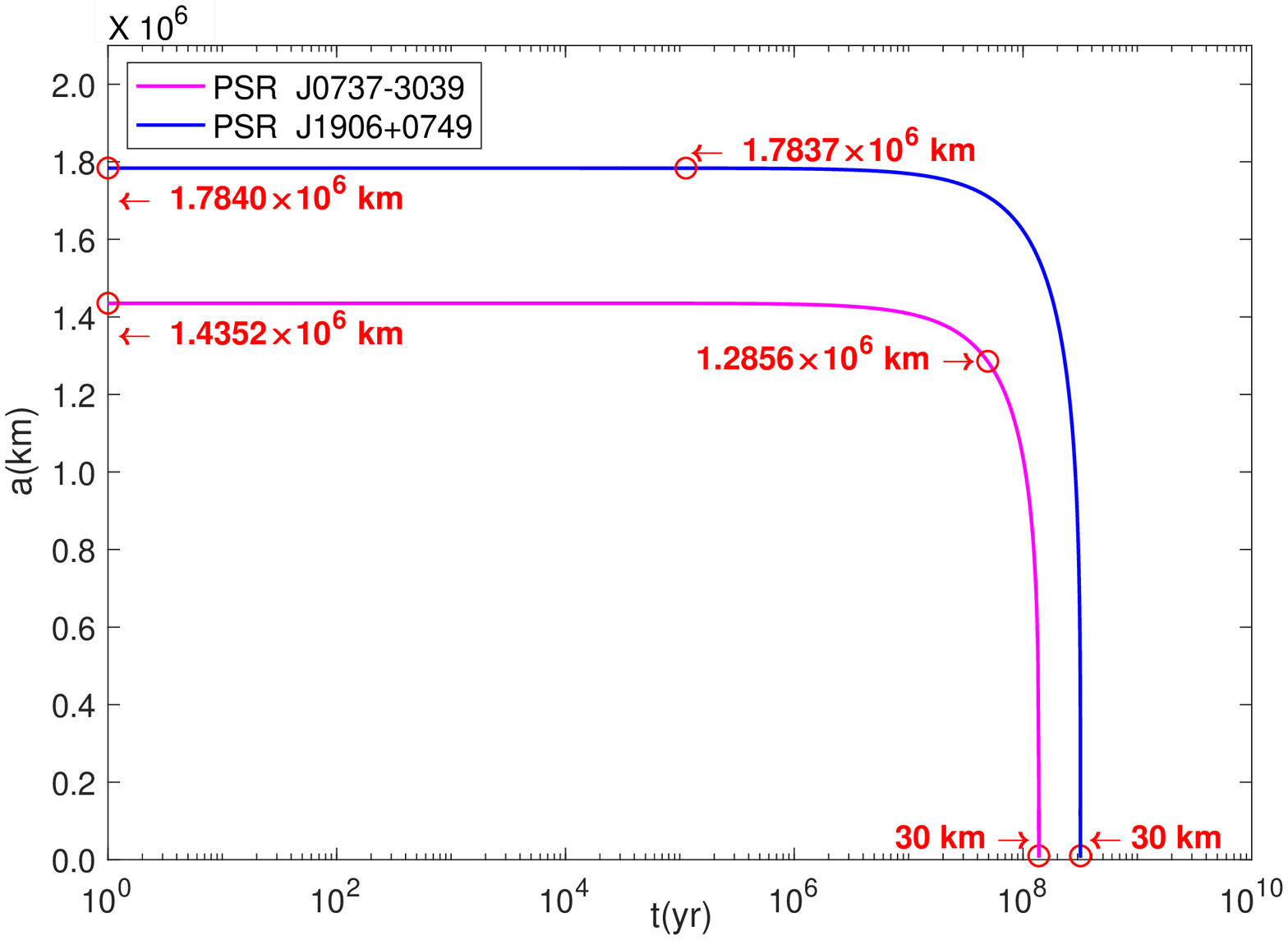}
\includegraphics[height=5cm]{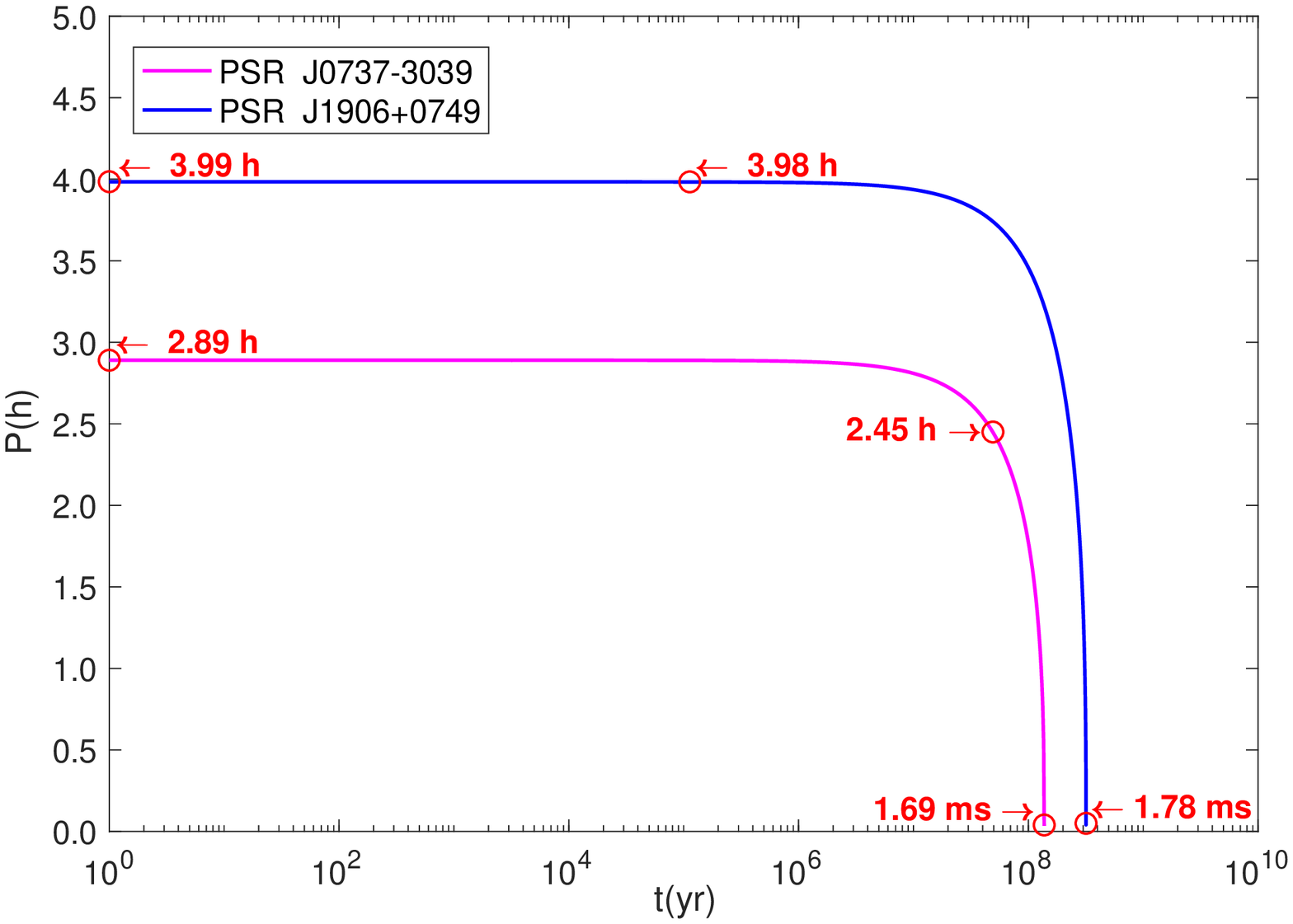}
\caption{The orbit evolution of double pulsar system PSR J0737-3039. As a comparison,  PSR J0737-3039 and PSR J1906+0749 are indicated.}
\label{f2}  
\end{figure*}

Based on the  magnetic dipole radiation model of pulsar (braking index n = 3) ~(\citealt{Lorimer+etal+2007, Shapiro+Teukolsky+1983}), we make a simple integral calculation for the pulsar spin-down model ($\dot{P_{s}} \varpropto P_{s}^{2-n}$), deriving the relationship among the true age (t), spin period ($P_{s}$), spin period derivative ($\dot{P_{s}}$), and initial spin period ($P_{s0}$) of the pulsar, which can be written as ~(\citealt{Lorimer+etal+2007, Shapiro+Teukolsky+1983, Zhang+etal+2016}):
\begin{equation}
\label{equ:7}
P_{s}^{2} = P_{s0}^{2} + 2t[P_{s} \dot{P_{s}}].
\end{equation}

Since the normal pulsar  (e.g., PSR J0737-3039B) has no experience of the accretion precess, its  initial spin period is  much smaller than its current spin period ($P_{s0} \ll P_{s}$), we can approximately deform Eq.(\ref {equ:7})  as:
\begin{equation}
\label{equ:8}
t \simeq  \frac{P_{s}} {2 \dot{P_{s}}}\equiv \tau,
\end{equation}
where $\tau$ is the current characteristic age of the pulsar, expressed as  $\tau \equiv P_{s}/(2 \dot{P_{s}})$. The above formula indicates that we can use the characteristic age of pulsar to approximately replace  its true age (when $P_{s0} \ll P_{s}$)~(\citealt{Lorimer+etal+2007, Camilo+etal+1994}).  
This means that the age of PSR J0737-3039 system is approximately equal to the current characteristic age of  PSR J0737-3039B.

Compared to PSR J0737-3039A that experienced the accretion recycled process, PSR J0737-3039B is a relatively young normal pulsar with the spin period of 2773.46 ms, with  the characteristic age of $4.92 \times10^ {7}$ yr~(\citealt{Burgay+etal+2003}).
Similar to Crab pulsar that has the constant magnetic field dipole radiation~(\citealt{Staelin+Reifenstein+1968}), we assume that the initial spin period of PSR J0737-3039B is 20 ms ~(\citealt{Zhang+etal+2016}), because the current spin period is much larger than the  birth spin period ($20 ~\rm ms \ll 2773.46 ~\rm ms$), therefore, consequently the PSR J0737-3039B true age ($t_{B}$) can be approximated by the characteristic age ($\tau_{B}$), namely:
 \begin{equation}
 \label{equ:9}
t_{B} \simeq \tau_{B}.
\end{equation}

Lorimer et al. (2007) pointed out that when the PSR J0737-3039A ceased to spin-up and the PSR J0737-3039B was birth occurred almost at the same time~(\citealt{Lorimer+etal+2007}), thus the spin-down age
of PSR J0737-3039A (time since the accretion induced spin-up stopped: $ t_ {A} $) can be approximately replaced by the true age of PSR J0737-3039B, that is:
 \begin{equation}
 \label{equ:10}
t_{A}  = t_{B} \simeq \tau_{B} .
\end{equation}

According to the standard DNS formation model~(\citealt{van den Heuvel+2007}), the first NS of the system was formed by a massive star undergoing main sequence evolution and supernova explosion. The first NS will
accrete matter from the  progenitor star of the companion star (B), which will  spin-up the period of the first NS; If the system survives in the second supernova explosion, the DNS system will be
born at this moment. Hence the true age of the DNS system is equal to the true age of the second NS, and the true age of PSR J0737-3039 system ($t_{AB}$):
 \begin{equation}
 \label{equ:11}
t_{AB} =  t_{B} \simeq \tau_{B} , 
\end{equation}
 from the above formula we can constrain the real age of  PSR J0737-3039 system to be $4.92 \times10^{7}$ yr, and the system will complete coalescence age (time from birth to merger) in approximately $1.38 \times10^{8}$ yr (the characteristic age of PSR J0737-3039B to be $4.92 \times10^{7}$ yr, the coalescence age of system to be $1.38 \times10^{8}$ yr).

We bring the age of PSR J0737-3039 system $t_{AB}$ into Eq.(\ref{equ:4}) and Eq.(\ref{equ:6}), and obtain the initial binary separation of the system to be  $1.44 \times10^{6}$ km,
corresponding to be initial orbital period of  2.89 hrs, which means that the binary separation and orbital period of system is only contracted  by  $0.15 \times10^{6}$ km and 0.44 hrs, respectively
(the present binary separation and orbital period  are $1.29 \times10^{6}$ km and 2.45 hrs, respectively). Next, based on the initial information of PSR J0737-3039 system, we simulated the
evolution of the orbital period and binary separation of PSR J0737-3039 within the complete coalescence age, as shown in Fig.\ref{f2}.


\subsection{The spin evolution of PSR J0737-3039A and PSR J0737-3039B in complete coalescence age}

\begin{figure}[htbp]
\centering
\includegraphics[scale=0.45]{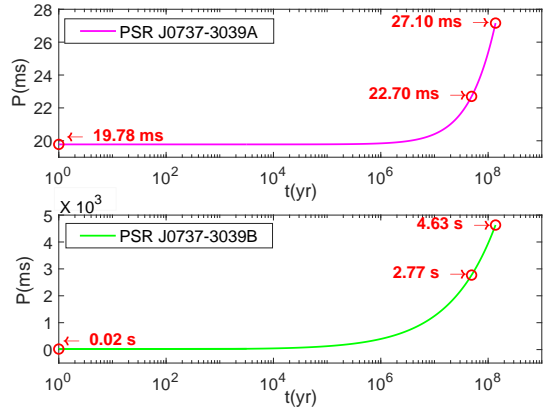}
\caption{Spin period evolutions of PSR J0737-3039A and PSR J0737-3039B. }
\label{f3}
\end{figure}

We bring the observational data of PSR J0737-3039A and $t_ {A}$ into Eq.(\ref {equ:7}), obtaining that the birth spin period is 19.78 ms when it starts to spin-down,  which means that its spin period is only increased by 2.92 ms (the present spin period is 22.70 ms). From the above, the initial spin period of PSR J0737-3039B is assumed to be similar to that of  the Crab-pulsar of 20 ms, compared to its current period of  2773.46 ms, thus  the spin period of PSR J0737-3039B evolves relatively quickly.

In order to simulate the evolution of the pulsar spin-down, we assume that the dipole magnetic field is invariant during the evolution, therefore we can transform Eq.(\ref {equ:7}) into:
\begin{equation}
\label{equ:12}
P_{s}^{2} = P_{0}^{2} + 2t[P_{n} \dot{P_{n}}],
\end{equation}
where $P_{n}$ and $\dot{P_{n}}$ are the current measured data of the pulsar spin period and spin period derivative, respectively.

According to Eq.(\ref {equ:12}), we simulated the spin period evolution of PSR J0737-3039A and PSR J0737-3039B within the complete merger age, and stopped the simulation when the evolution time going to $ 1.38 \times10 ^ {8} $ yr.
Our simulation found that:  the spin period of PSR J0737-3039A and PSR J0737-3039B are 27.10 ms and 4.63 s, respectively, when the system was merged. Based on the simulation results, we plot the spin period   evolution   of  PSR J0737-3039A and PSR J0737-3039B in Fig.\ref {f3}.

\begin{figure}[htbp]
\centering
\includegraphics[scale=0.45]{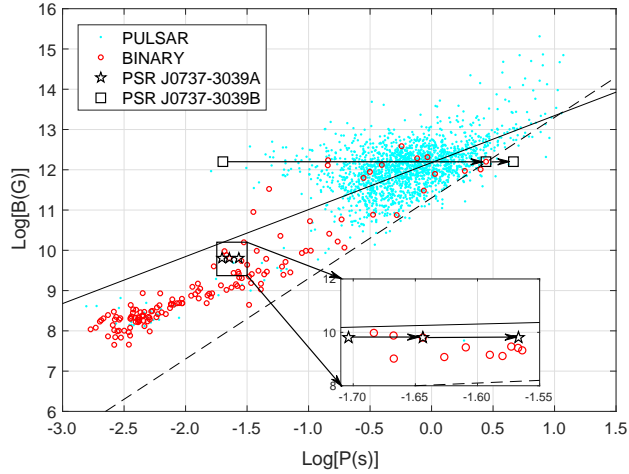}
\caption{The evolution track of PSR J0737-3039A and PSR J0737-3039B along the $B-P$ diagram (data from ATNF pulsar catalog~(\citealt{Manchester+etal+2005}). The cyan dots, red circles, black lines, black dotted lines,
 pentagrams and squares are  represent normal pulsars, pulsed binary, acceleration lines, death lines, PSR J0737-3039A and PSR J0737-3039B, respectively.}
\label{f4}
\end{figure}

In order to acquire a complete scenario of  the evolution of both pulsars in the diagram of spin period and   magnetic field (the magnetic fields of PSR J0737-3039A and PSR J0737-3039B are $6.4 \times10^{9}$ G and $1.59\times10^{12}$ G,
respectively~(\citealt{Burgay+etal+2003, Lyne+etal+2004}), we plot their complete simulation results  in Fig.\ref{f4}.

In the Figure 4, the small box  is a close-up of the evolution track of PSR J0737-3039A, and the arrows indicate the direction of evolution (the evolution from left to right  means that from birth to merger). It can be seen from the figure that from the system birth to merger, PSR J0737-3039A would be  always  seen by the radio telescope, while PSR J0737-3039B will cross the death line of pulsar  and may be not observed ~(\citealt{Hulse+van den Heuvel+1991}).

\section {The results and Discussions }
\label{sect:result}

\begin{figure}[htbp]
\centering
\includegraphics[scale=0.52]{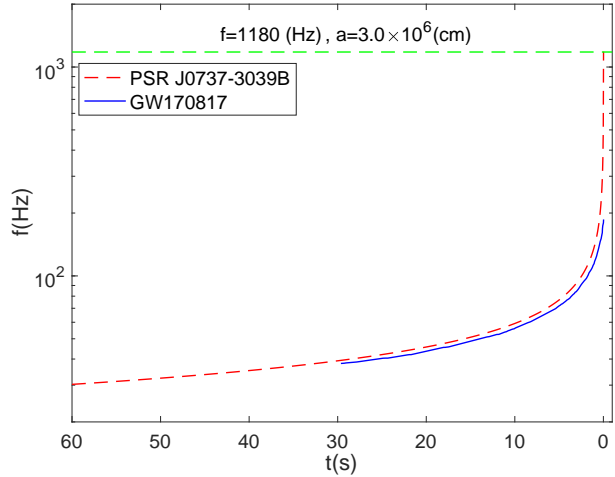}
\caption{ The evolution of GW frequency at the last minute of DNS merge, where the dashed (solid) curve  represents the case of  PSR J0737-3039 (GW170817).} \label{f5}
\end{figure}

In this paper, based on the observational data of the double pulsar system PSR J0737-3039, we simulated its orbital evolution induced by the GW radiation, and found that the system will merge after $8.83 \times 10^{7}$ yr. By the magnetic dipole model of pulsar, we obtained  that  PSR J0737-3039 system exists about  $4.92 \times10^ {7}$ yr, thus the complete coalescence age (from its birth to merge) of  PSR J0737-3039 system should be be $1.38 \times10^ {8}$ yr.
Next, we simulated the complete orbital evolution of PSR J0737-3039 system and obtained the  initial orbital period and radius as 2.89 hrs and 1.44 km, respectively. In addition,  the GW frequency generated within last minute before the system merge ranges  from 20 Hz to 1180 Hz, and the corresponding binary separation decays from 442 km to 30 km. We compared  the GW frequency evolution  of PSR J10737-3039 with that of  the  observation data of GW170817 by LIGO, as shown in Fig.\ref{f5}, and found that both curves have no much difference, and a meager bias is on account of the mass difference of both systems (the primary NS and companion NS masses of GW170817 are: $1.46 _{-0.10}^{+0.12} M_{\odot}$  and $1.27_{-0.09}^{+0.09}$, respectively ~(\citealt{Abbott+etal+2018}).

Furthermore,  we calculated  the  spin period evolution of PSR J0737-3039A and PSR J0737-3039B within the complete merger age, and obtained that their spin periods  are 27.10 ms and 4.63 s at coalescence.
Through the evolution trajectories of PSR J0737-3039A and PSR J0737-3039B in the pulsar magnetic field and spin period  $B-P$ diagram, we concluded that  PSR J0737-3039A can always be observed by radio telescopes, but  PSR J0737-3039B will cross the death line of radio pulsar and could not  be observed in general.

It is remarked that we employed the point masses  of both NSs to perform the simulation. However,   Kuznetsov et al. (1998) studied the orbital evolution of DNS by the GW radiation while the internal structure of NSs have been taken into account, and pointed out that there exists little effect on their merging  time (only deviation of  10.5 ms)~(\citealt{Kuznetsov+etal+1998}), which is much
less than the merging time of  30 s as  observed in  GW170817 by LIGO.

\begin{acknowledgements}
This research is supported by the National Natural Science Foundation of China NSFC (11988101, 11773005, U1631236, 11703001, U1731238, U1938117, 11725313, 11721303), the International Partnership Program of Chinese Academy of Sciences grant No. 114A11KYSB20160008, the National Science Foundation of China No. 11721303, and the National Key R\&D Program of China No. 2016YFA0400702. And, this work is also supported by the National Basic Research
Program (973 Program) (no. 2015CB857100), National Key R\&D Program of China (no. 2017YFA0402600), and the Guizhou Provincial Science and Technology Foundation (Grant no. [2020]1Y019).
\end{acknowledgements}


\label{lastpage}

\end{document}